# Charge transfer tuning in TiO$_2$ hybrid nanostructures with acceptor-acceptor systems

K. Pilarczyk,[a, b] K. Lewandowska,[c] K. Mech,[b] M. Kawa,[b, d] M. Gajewska,[b] B. Barszcz,[c] A. Bogucki,[c]

A. Podborska[b*] and K. Szaciłowski[b*]

a.  AGH University of Science and Technology, Faculty of Physics and Applied Computer Science, al. A. Mickiewicza 30, 30-059 Kraków, Poland.

b.  AGH University of Science and Technology, Academic Centre for Materials and Nanotechnology, al. A. Mickiewicza 30, 30-059 Kraków, Poland.

c.  Polish Academy of Science, ul. Smoluchowskiego 17, 60-179 Poznań, Poland.

d.  AGH University of Science and Technology, Faculty of Energy and Fuels, al. A. Mickiewicza 30, 30-059 Kraków, Poland.

† e-mail: podborsk@agh.edu.pl, szacilow@agh.edu.pl

An interesting interplay between two different modifiers and the surface of titanium dioxide leads to a significant change in photoelectrochemical properties of the designed hybrid materials. The semiconductor is photosensitized by one of the counterparts and exhibits the photoelectrochemical photocurrent switching effect thanks to interactions with graphene oxide – the second modifier mediates charge transfer processes in the system, allowing us to design the materials response at the molecular level. Based on the selection of molecular counterpart we may affect the behaviour of hybrids upon light irradiation in a different manner, which may be useful for the applications in photovoltaics, optoelectronics and photocatalysis. Here we focus particularly on the nanocomposites made of titanium dioxide with graphene oxide combined with either 2,3,5,6-tetrachlorobenzoquinone or 2,3-dichloro-5,6-dihydroxybenzoquinone – for these two materials we observed a major change in the charge transfer processes occurring in the system.

**Introduction**

The research efforts focused on the synthesis and the characterisation of new functional materials are dictated by the growing interest in alternative approaches to the design and fabrication of computing devices,[1-5] photovoltaic components,[6,7] chemo- and biosensors,[8-11] etc. Although, these systems will probably not threaten the well-established position of classical, silicon-based electronic elements, they can serve as a good supplement, enhancing the possible interactions of devices with ions, light and other information carriers.

Among others, molecular materials provide the necessary variety of interactions which can be utilised to make the designed components more versatile, because properties of these materials can be fine-tuned by subtle alterations of the modifer.[12] Therefore, we can observe intensification of work in the fields such as molecular electronics and spintronics,[13-16] molecular-based sensing,[17-19] molecular-based photovoltaics[20-23] and many others. In other words, we may expect that this trend will continue giving rise to new applications in which silicon could not be used. Supramolecular entities and carbon nanostructures are considered to be a good linker between the molecular world and the realm of semiconducting materials (in the form of nanoparticles, thin films, etc.). Many scientific groups conduct research on the use of carbon nanotubes, graphene, fullerenes and other nanoforms of carbon as modifiers for semiconductors.[24-28] One of the most intriguing representative of this group is graphene oxide (GO), which due to the presence of various functional groups (hydroxyl, epoxy, carboxylic, carbonyl) as well as pristine graphene domains is a versatile playground for covalent and noncovalent modifications.[26,29] Moreover it can be fairly easy obtained from commonly used reagents in several different oxidation procedures (*e.g.* derived from the original Hummers method).[30]

There are numerous papers dealing with the semiconducting and metallic nanoparticles interacting with graphene oxide. These hybrids are obtained by the decoration of GO sheets with inorganic nanoparticles and they are considered for applications in electrochemical sensing devices, catalysis, and fuel cells. These functional metal or metal oxide/GO nanocomposites can be prepared using different physical or chemical approaches: a physical attachment approach, an in-situ chemical reduction process, electrochemical synthetic processes, impregnation processes, a self-assembly approach, ultrasonic spray pyrolysis and others. Numerous GO-based nanocomposites with metal nanoparticles and oxide nanoparticles have been reported recently.[31, 32] The authors focus on different aspects of hybrid materials properties, but all of them emphasise that due to GO characteristics, it may provide a good platform for tunable systems for applications in electronics and photovoltaics. At the same time, it is tempting to take an advantage of intrinsic properties of GO and to combine it with molecular modifiers.[33, 34] The nature of interactions in such binary modifier influences the optical and electronic properties of semiconductor particles via the change in the character of the surface states and by the incorporation of additional energy levels. Altogether, this kind of hybrids could exhibit complex emergent features which might be useful in prototypic photomodulated electronic devices.

In this paper we investigate interactions between graphene oxide and two benzoquinone derivatives (namely 2,3,5,6-tetra-chlorobenzoquinone and 2,3-dichloro-5,6-dihydroxybenzoquinone) with titanium dioxide. We are particularly interested in the tuning of photoelectrochemical properties of such hybrid materials associated with the interplay between two electron acceptors coupled with semiconducting nanoparticles. The binary systems (i.e. semiconducting nanostructures with either graphene oxide or a molecular modifier) are well described in the literature, whereas ternary nanocomposites are still underrepresented. The introduction of a third player to the system (e.g. the molecular modifier) may facilitate the desirable charge transfer and energy transfer processes, which is of crucial significance for the applications in photovoltaics, optoelectronics and photocatalysis.

**Experimental**

Graphene oxide was synthesized from natural graphite powder by the modified Hummers method.[35, 36] Graphite powder (3 g) and $KNO_3$ (3 g) were added to concentrated $H_2SO_4$ (90 ml) and the mixture was stirred in ice bath. Small portions of the oxidizing agent $KMnO_4$ (9 g) were added slowly in order to keep the suspension temperature below 2°C. The reaction mixture was maintained at approx. 0°C and it was vigorously stirred for 15 min resulting in the increase of temperature to 35°C. The stirring was continued for 7 h.

Afterwards, 90 ml of distilled water was slowly added and the mixture temperature increased to 80°C. After 15 min the suspension was cooled to room temperature and the reaction was quenched by the addition of 12 ml of $H_2O_2$ (30%). In the first step of the purification process the diluted (with 250 ml of distilled water) supernatant was poured off. The precipitant was rinsed two times with 250 ml of distilled water and the supernatant was removed. Then the product was centrifuged in 1 mol/$dm^3$ HCl aqueous solution at 6000 rpm for 3 min three times (to remove manganese compounds) with following decantation. Afterwards, the process was repeated four times with distilled water. The final product was dried at room temperature.

The suspension of GO in acetonitrile (ACN) was prepared by sonication of dispersed GO powder (20 mg) in ACN (200 ml). The process was repeated three times for 30 min. The ACN solutions of quinone modifiers were prepared by the dissolution of 1,4-benzoquinones derivatives: 2,3,5,6-tetrachloro-1,4-benzoquinon (CLA, 12.29 mg) and 2,5-dihydroksy-3,6-dichloro-1,4-benzoquinon (KCLA, 10.45 mg) in acetonitrile (50 ml). In the next step 25 ml of the quinone solutions were mixed with 10 ml of either ACN or GO suspension yielding 4 samples altogether (CLA, KCLA, CLA/GO and KCLA/GO). For comparison reasons a sample containing only GO was prepared – to 10 ml of GO suspension 25 ml of ACN was added.

The hybrid materials (with $TiO_2$ and CdS) were prepared by the impregnation of fine powder with four aforementioned solutions. Evonik P25 (anatase/rutile blend of approx. 80:20 ratio)[37] and a hexagonal CdS (POCH, Poland) were used in this study. In each case 0.1 g of semiconductor was mixed with 10 ml of the solution and sonicated for 20 min. The resulting precipitate was centrifuged at 5000 rpm for 5 min. The process was repeated 4 times with the addition of 10 ml of ACN in each case.

The modifiers solutions were investigated with the use of UV-Vis spectrophotometry and spectrofluorometry. In the first case all the samples were prepared by the dilution of the initial solution (200 µl) with acetonitrile (3ml). The control system containing only GO was mixed with ACN in 1:1 v/v ratio. The spectra were recorded with the use of Agilent 8454 UV-Vis spectrophotometer.

The fluorimetry measurements were carried out on a Fluoromax 4P spectrofluorometer (Horiba Jobin Yvon, France). The samples were prepared as follows: 0.5 ml of the initial solution was mixed with 3 ml of ACN and the spectra were recorded with the excitation wavelengths equal to 300 nm (for KCLA and KCLA/GO) and 290 nm (for CLA and CLA/GO) which corresponds to the peaks in the absorption spectra.

ATR-IR spectra were recorded for the samples drop-casted onto the Ge crystal (for suspensions) or deposited onto diamond crystal (in the case of hybrids) with the use of FT-IR Bruker TENSOR II spectrometer within the range of 600-4000 cm$^{-1}$ at the room temperature. Raman scattering spectra of the investigated systems were recorded using LabRAM HR 800 spectrophotometer (Horiba Jobin Yvon, France) with the excitation wavelength equal to 633 nm. In all the cases the power of the laser beam was lower than 1 mW with the power density of approximately $3 \cdot 10^8$ mW·cm$^{-2}$. Such a low power density was necessary to avoid decomposition of the samples.

In order to determine the optimal structure of the compounds and to interpret the experimental results of IR absorption and Raman scattering measurements the quantum chemical calculations were carried out (Scheme S1, Table S1-S2). The molecular geometries were optimized using the Density Functional Theory (DFT) method with B3LYP hybrid functional and 6-31G basis set. The calculations of normal mode frequencies and intensities were also performed. All calculations were done using Gaussian 03 package.[38] The GaussView program was used to establish the initial geometry of investigated molecules and for the visualisation of the normal modes.

The structure and morphology of the hybrid materials (containing TiO$_2$ and CdS modified with CLA, KCLA, CLA/GO and KCLA/GO) were investigated with the use of powder X-ray diffractometry (XRD) and the electron microscopy imaging (in both scanning, SEM and transmission, TEM modes). XRD patterns were recorded on Panalytical Empyrean diffractometer with Cu K$_\alpha$ line within the 2θ range between 5° and 80°. SEM study was carried out on FEI Versa 3D (FEG), operating in a low vacuum conditions, using Helix Low-Vacuum SE Detector. TEM imaging was performed using FEI Tecnai TF20 X-TWIN (FEG) microscope, at the accelerating voltage of 200 kV.

The optical properties of nanocomposites were determined based on the diffusive reflectance spectrophotometry within the range of 200–2200 nm. The measurements were carried out using Lambda 950 (Perkin Elmer, USA) spectrophotometer equipped with a 150 mm integration sphere. Each sample was dispersed in spectrally pure BaSO$_4$. The pressed BaSO$_4$ pellet was used as a reference.

In order to perform the photoelectrochemical characterization of the samples the modified materials were dispersed in a small volume of distilled water and drop-casted onto the surface of ITO@PET. The measurements were done using a photoelectric spectrometer

(Instytut Fotonowy, Poland) composed of stabilized 150W xenon arc lamp, monochromator and coupled with the SP-300 potentiostat (Bio-Logic, France). The photocurrents were recorded using a classical three electrode setup with a platinum wire counter electrode and an Ag/AgCl sat. reference electrode. The electrolyte solution was 0.1 mol/dm$^3$ KNO$_3$ saturated with either oxygen or argon.

The same sample preparation method as well as electrochemical system were used for electrochemical impedance spectroscopy (EIS) characterisation of synthesized materials. EIS spectra were recorded within potential range from 0.6 V to -0.2 V, with a potential step of 0.05 V. The electrode was conditioned at corresponding potential for 5 s before each measurement. Several potentials spectra were recorded starting from 200 kHz to 1 Hz (with 10 points per decade) for potential amplitude of 0.08 V. EIS analysis was performed in the argon saturated (purged for 5 min) electrolyte containing 0.1 mol/dm$^3$ KNO$_3$.

If not stated otherwise, all the chemicals were supplied by Sigma-Aldrich.

**Results and discussion**

In order to determine possible interactions between both molecular counterparts UV-Vis absorption spectra were recorded (Figure S1). The results obtained for both dyads are comparable to the data acquired for solutions of respective quinone derivatives. The positive shift in intensities (within the whole examined range) can be attributed to the absorption of graphene oxide (it is important to note, that the concentration of the sample containing exclusively GO was higher for readability reasons). Particularly, no new absorption bands were observed and no shifts in existing peaks (associated with the absorption of quinone derivatives) energies were detected. That may indicate that no new interactions (*e.g.* charge transfer processes) between molecular entities emerge. The same conclusion is also valid for the varying concentration ratios between both counterparts.

It is noteworthy that the spectrum recorded for the graphene oxide dispersion exhibits only a weak absorption band at 230 nm, which is distinctive for GO. At the same time, the presence of shoulders at higher wavelengths indicate that a partial reduction of the material took place – probably during the sonication of the mixture.[39, 40]

The additional evidence of no relevant interactions between two counterparts of the investigated systems is provided by the results of spectrofluorometry measurements. Again, the excitation at the wavelengths corresponding to the absorption peaks of quinone derivatives gives the spectra which are very similar (with only a minor distortion in the case of KCLA/GO system) in pairs with and without the addition of GO (Figure S2). Neither static nor dynamic quenching processes were identified. The recurring peaks (marked with arrows) were assigned as the first order Raman peaks of the solvent.

In order to verify the possibility of new bonds formation between GO and quinone derivatives we measured IR and Raman spectra of the samples. In the former case, it is evident that no new signals appear and it may be interpreted as an indirect proof that no strong interactions emerge in the mixtures of CLA/GO and KCLA/GO (Figure 1). In the IR spectrum of GO we can observe a pattern which is commonly reported for this system.[41, 42]

In the case of CLA the comparison of the experimental data with the results of DFT modeling (which fits the measured spectrum relatively well – please compare with Table S1) reveals the origin of several absorption bands. The absence of some bands related to the graphene oxide in the CLA/GO mixture may be explained by a low concentration of GO and relatively intensive signals from the benzoquinone derivative.

As for KCLA and its dyad with GO the experimental spectra are also relatively well approximated by the DFT calculations results (please refer to ESI – Table S2), nonetheless some significant discrepancies may be pointed out. The most substantial shifts occur in the range of 1200-1400 $cm^{-1}$ – where C-O-H bending vibrations are observed – which may be attributed to the strong interactions of hydroxyl groups in the solid phase. That contribution was not taken into account in the calculations, thus observed differences emerged. It should be noted, that in this case a major change in relative intensities of several peaks is noticeable in KCLA/GO system – mainly in the region around 1667 $cm^{-1}$ (1681 $cm^{-1}$ for CLA) and 3300 $cm^{-1}$. These bands are assigned to C=O and O-H vibrations, respectively. The decrease in their intensity indicates the formation of hydrogen bonds between GO and quinonic modifiers.

The Raman spectra of both dyads are dominated by so called D (1340 $cm^{-1}$) and G (1594 $cm^{-1}$) bands attributed to the presence of graphene oxide (Figure 2). The position of peaks is consistent with values found in the literature and the slight shift (with respect to graphene) may be connected with the decrease in the size of in-plane $sp^2$ domains and the presence of isolated double bonds that resonate at the higher frequencies than the position of the G band of graphite.[43]

A partial agreement with the calculated spectra was achieved in case of both quinone derivatives. The detailed description of individual bands is presented in the Electronic

Supplementary Information (Table S1-S2). The most important information derived from the results obtained for CLA/GO system is an additional evidence of no interactions between both components – the peaks associated with CLA are fully reproduced in the case of CLA/GO sample.

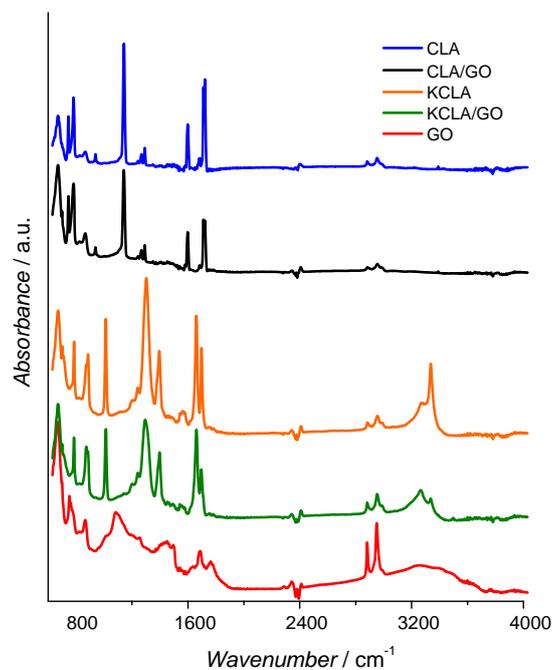

Figure 1. The ATR-IR spectra of quinone derivatives, dyads with graphene oxide and GO drop-casted onto Ge crystal from acetonitrile suspensions. The results for CLA and KCLA are complemented with DFT calculations results (Scheme S1 and Table S1-S2).

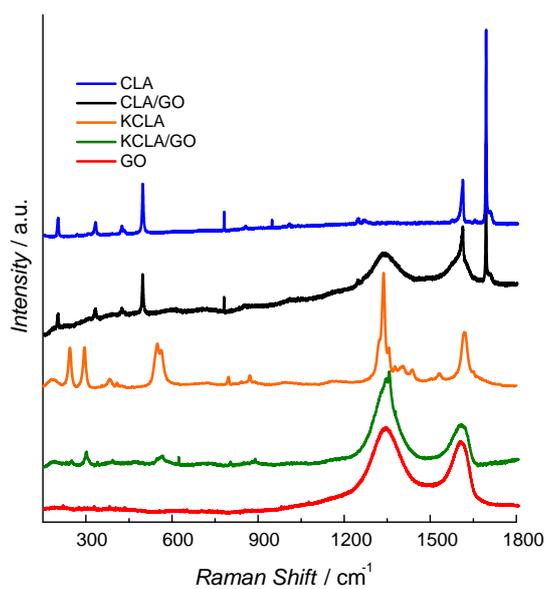

Figure 2. The Raman spectra of quinone derivatives and dyads with graphene oxide. The results for CLA and KCLA are complemented with DFT calculations results (Scheme S1 and Table S1-S2).

At the same time, the data recorded for KCLA and its assembly with GO becomes devoid of one peak attributed to the quinone below 300 cm$^{-1}$. One possible explanation of this change is a stronger interaction between GO and KCLA, which is partially confirmed by a higher D/G peak area ratio, which is equal for KCLA/GO to 2.54 (whereas it is equal to 1.63 in the case of CLA/GO and 2.12 for unmodified graphene oxide). That may indicate that significant interactions occur between π-electron system of chloranilic acid and graphene oxide conjugated system, which could be also connected with a redox reaction occurring in the sample (please refer also to the comments in ESI and Figures S3-S5).

The next phase of the experimental work was focused on the determination of an influence of described acceptor-acceptor binary modifiers on the photoelectrochemical properties of wide-bandgap semiconductors. We were particularly interested in the changes of a photocurrent response of the obtained hybrid systems with oxides and sulphides of transition metals (in this case TiO$_2$ and CdS).

Cadmium sulphide has been selected for the comparison due to two fundamental reasons: (i) the difference in the electronic configuration of metal ions ($d^{10}$ vs. $d^0$ in TiO$_2$) which, in the case of CdS excludes the possibility of the charge transfer complex formation with semiquinone-type ligands and (ii) the common use of CdS in works on optoelectronic logic gates.

The electron micrography (both SEM and TEM) clearly indicates differences in interactions between graphene oxide and TiO$_2$ in the absence and the in presence of the KCLA modifier. Titanium dioxide nanoparticles are usually evenly distributed on graphene oxide sheets (Figures 3a and 3b). Upon the introduction of KCLA the interplay between components leads to the aggregation of TiO$_2$ nanoparticles and the adsorption of these aggregates at the graphene oxide sheets (Figure 3c and 3d). It may be concluded that the surface ligands (KCLA) facilitate the aggregation of TiO$_2$ and that these ligands promote the interactions with graphene oxide aromatic regions, therefore large areas of GO flakes are not covered by modified TiO$_2$ nanoparticles – TiO$_2$ itself should favour polar, highly oxidized regions.

Along with the morphology studies the structure of crystallites was examined. The XRD diffractograms (Figure S6 in ESI) revealed the presence of both rutile and anatase phases in TiO$_2$ sample and a hexagonal polymorph in the case of CdS. Moreover, authors also verified the influence of GO introduction on the crystal structure. It was concluded that no significant changes appear in the patterns upon the modification.

In order to determine particular interactions between graphene oxide, the molecular modifier and semiconductor nanoparticles the UV-Vis diffused reflectance spectroscopy was used. The spectra were converted using the Kubelka-Munk transformation. Then, the Tauc plots were prepared to check if the bandgap width is affected by the addition of molecular modifiers. The obtained results are summarised in Figure 4 for TiO$_2$-based materials and in Figure S7 for CdS-based systems.

The analysis of the results obtained for titanium dioxide and cadmium sulphide (cf. ESI, Figure S7) reveals some interactions between semiconductors and dyadic modifiers. Significant changes may be noticed in the system containing KCLA which agrees with the conclusion drawn before on the reactivity of KCLA. One of the most important changes can be found in KCLA/TiO$_2$ and KCLA/GO/TiO$_2$ spectra where an additional charge-transfer band appears at approx. 2.25 eV (Figure 4).

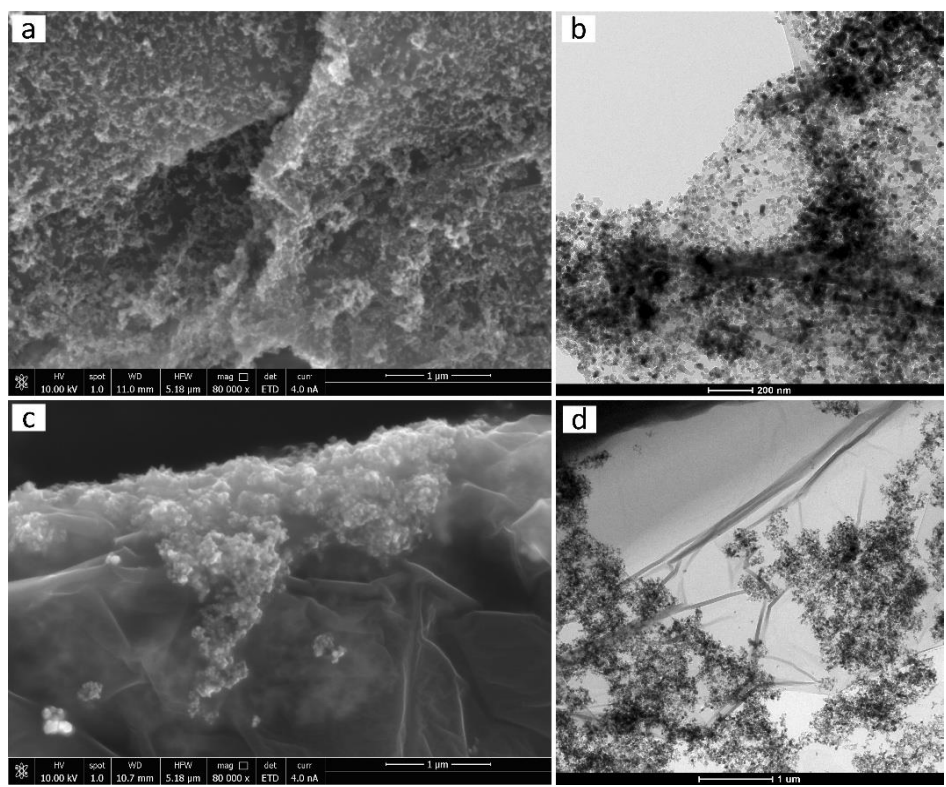

Figure 3. Scanning electron images (a, c) and transmission electron micrographs of GO/TiO$_2$ (a,b) and GO/KCLA/TiO$_2$ (c, d).

This new interaction results from direct covalent interactions between quinone and surface titanium ions, most probably mediated by hydroxyl and carbonyl groups. The CLA molecule may bound relatively weaker as no chelating moiety is present. Thus, we observe only weak absorption (at ca. 2.5-3.0 eV) which may be associated with the formation of a charge-transfer complex. Anyway, in both cases a complex mechanism, involving reduction of quinone to semiquinone anion radical followed by the chemisorption must occur. A closely related process was already observed for tetracyanoquinodimethane and related molecules.[44-47] In this case, however, the final product is formed without the significant rearrangement of molecular species. The high stability of these materials is substantiated by the stabilization of semiquinone radicals by high oxidation state (here $d^0$) metal centres.[48]

DFT calculations indicate that semiquinone anion radical of KCLA binds to titanium centres on the surface of TiO$_2$ with both phenolic and quinonic groups. Thus, the formed chelating structure is stable and the unpaired electron, occupying the semioccupied molecular orbital (SOMO) is delocalized over aromatic and inorganic parts (Figure 5). Furthermore, the contours of the highest occupied orbital (HOMO) and the lowest unoccupied orbital (LUMO) indicate strong charge-transfer interaction between organic and inorganic moieties, like in the catecholate and salicylate complexes assembled at TiO$_2$ surfaces.[14, 49]

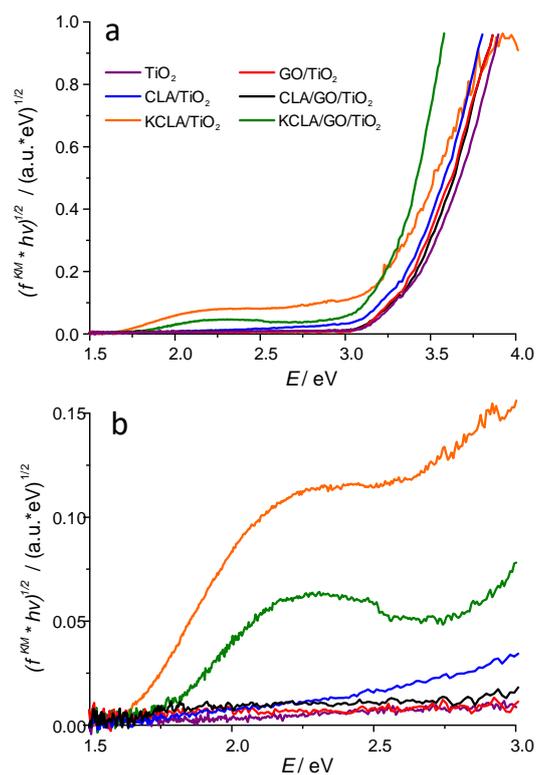

Figure 4. The Tauc plots for titanium dioxide and the hybrid materials containing TiO$_2$ (a) and an expanded view of the low energy region in which charge transfer bands are visible (b).

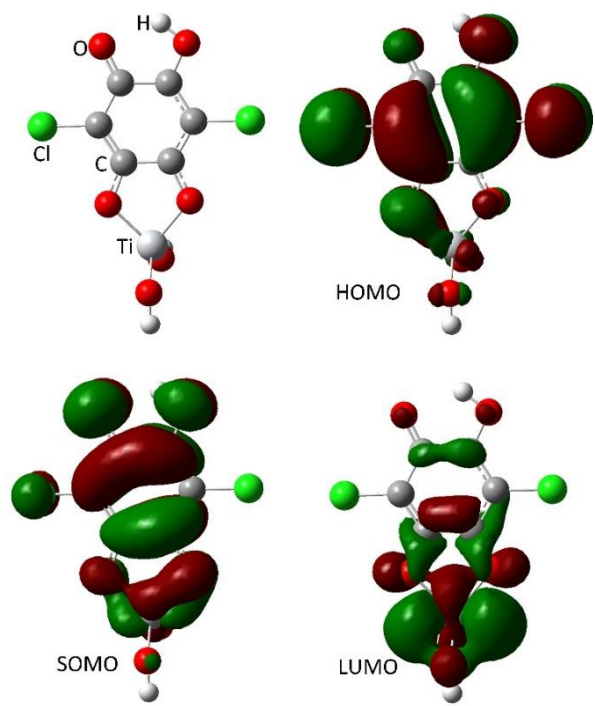

Figure 5. The structure of a minimalistic model used for the quantum-chemical analysis for interactions between the semiquinone radical anion and titanium dioxide along with the contours of frontier orbitals.

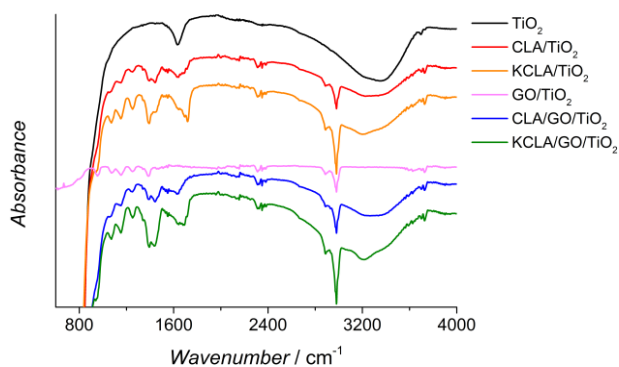

Figure 6. The ATR-IR spectra of $TiO_2$ hybrids with quinone derivatives, dyads with graphene oxide and GO deposited onto diamond crystal.

Additional evidence for the complex nature of interactions between the constituents of $TiO_2$-based hybrids, involving mainly oxygen-rich functional groups (particularly hydroxyl groups) is provided by the ATR-IR spectra of nanocomposites (Figure 6). The analysis indicates some deformations within the region in which signals originating from –OH are typically recorded (i.e. above 3200 $cm^{-1}$) and in the bands at approximately 1600 $cm^{-1}$. Interestingly, the spectrum obtained for $GO/TiO_2$ system lacks of these distinctive peaks (the result is fully reproducible), which could be explained by a strong coupling between bare $TiO_2$ nanoparticles and the surface of graphene oxide – this is consistent with the SEM images. The results obtained for CdS-based nanocomposites are presented in ESI (Figure S8).

Interactions of titanium dioxide with quinone electron acceptors (especially when it leads to formation of semiquinone anion radicals) should result in cathodic shift of the conduction band edge if the Fermi level pinning process is assumed. In other words, these interactions should induce a decrease in electron concentration in the n-type semiconductor. In order to address this issue the Mott-Schottky analysis of $TiO_2$ and its hybrids was performed, which showed an impact of $TiO_2$ modification by KCLA, CLA or GO on the flat-band potential (i.e. a parameter closely related to the conduction band edge potential). The flat-band potential value in the discussed case was determined based on the Mott-Schottky plot (Figure 7). The impedance spectra were fitted using a basic equivalent circuit composed of a resistor (R) and a constant phase element (CPE). The CPE was used in order to replace the standard capacitance enabling application of the method for porous materials. The impedance of the CPE is then expressed by equation 1.

$$Z_{CPE} = \frac{1}{Q(i\omega)^{\alpha}} \quad (1)$$

where $Q$ stands for charge, $j$ is the imaginary unit, $\omega$ depicts angular frequency and $\alpha$ is a constant (for capacitors $\alpha=1$). The interfacial capacitance (C), strongly dependent on the applied voltage value (E) is described by the Mott-Schottky equation:

$$\frac{1}{C^2} = \frac{2}{\varepsilon\varepsilon_0 A^2 e N_D}(E - E_{VB} - \frac{k_B T}{e}) \quad (2)$$

where $A$ is the interfacial area, $N_D$ is the number of donors, $k_B$ is Boltzmann's constant, $T$ stands for the absolute temperature, $\varepsilon$ is the dielectric constant of layer, $\varepsilon_0$ is the vacuum permittivity and $e$ is the elementary charge.

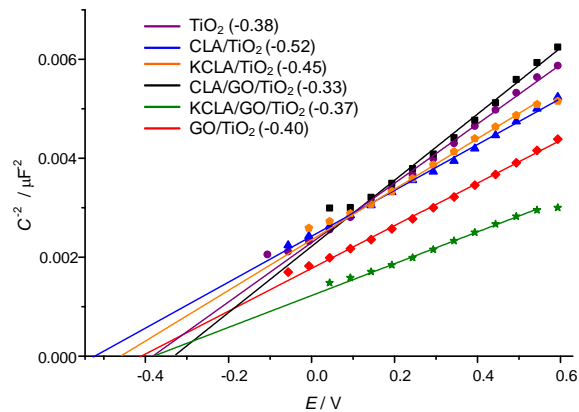

Figure 7. The Mott-Schottky plots for various TiO$_2$-based materials. Numbers in brackets indicate the flat-band potentials vs. Ag/AgCl (sat. KCl).

Figure 7 presents plots which are characteristic for n-type semiconductors. The determined values of the flat-band potentials indicated some differences between the hybrid materials of varying composition. The most significant changes, resulting in a shift of $E_{VB}$ towards more electronegative potentials were observed for TiO$_2$ modified with quinone derivatives. This observation is fully consistent with the aforementioned mechanism describing the reaction between titanium dioxide and quinone electron acceptors. The modification of TiO$_2$ with GO resulted only in a slight shift of $E_{VB}$ by 0.02 V in the cathodic direction, which may suggest that similar type of interaction with the surface occurs for both quinone functional groups and the oxidized edges of graphene oxide nanoflakes. The opposite effect represented by the anodic shifts of $E_{VB}$ was observed for CLA/GO/TiO$_2$ as well as KCLA/GO/TiO$_2$ hybrid systems.

The obtained results, which are in a good agreement with the photocurrent action spectra presented in Figure 8, are apparently counterintuitive. It should be realised, however, that the used method probes the *electrostatic state of the surface*. The studied material is characterised by a layered structure. TiO$_2$ bounds the semiquinone radical anions which constitute the layer of the negative charge. At the top of this layer graphene oxide is adsorbed, thus it becomes positively charged due to electrostatic interactions. Therefore, stronger electron acceptors should induce stronger anodic shifts of the flat band potential in the presence of graphene oxide adlayer, as observed in the experiment (cf. Figure 7). The modification of TiO$_2$ with KCLA resulted in the intensification of cathodic photocurrents, whereas addition of KCLA/GO dyad promotes anodic photocurrents. It should be pointed out that more noticeable effect of the modification (in terms of the $E_{VB}$ shift) is visible for systems containing CLA.

Knowing that the titanium dioxide strongly bounds the modifiers to its surface we investigated the photocurrent response of hybrid systems containing TiO$_2$ (Figure 8). As expected, the most substantial changes were observed for KCLA-containing materials, but some new phenomena emerged in the case of the GO addition. The most prominent alteration was noticed when the KCLA/TiO$_2$ sample was compared with the KCLA/GO/TiO$_2$ hybrid.

The spectrum obtained for unmodified titanium dioxide is a typical spectrum obtained for an n-type semiconductor. Anodic photocurrents are recorded in the whole absorption range of the material and within the whole potential range (Figure 8a). A small decrease of photocurrent intensity with a drop in the photoelectrode potential is fully consistent with the Butler equation.49 In the case of GO/TiO$_2$ material (Figure 8b) only a slight increase of the photocurrent intensity (as compared with neat TiO$_2$) at lower potentials was noticed. This may indicate the existence on an another electron transfer pathway from the conduction band of semiconductor to the electrode with empty states of graphene oxide involved (cf. Figure 9c). The adsorption of KCLA (to some extent also CLA) results in the change in the photocurrent polarity and also some minor photosensitization within the visible light region (Figure 8c) .It is noteworthy that the cathodic signal dominates in the whole range of investigated potentials irrespectively of the presence of molecular oxygen (Figure S9). These variations are related with an electron acceptor character of surface molecules. Furthermore, upon formation of semiquinone (cf. Figure 5) the emergence of additional light absorption processes (cf. Figure 4b) is possible. When both modifiers are present two phenomena become superimposed (Figure 8d). In such a case the photosensitization towards visible light is observed. At the same time, amplified anodic photocurrents strongly influence the photocurrent characteristics and the photocurrent switching effect appears. A complex wavelength-dependent photocurrent switching is observed in the potential range of 0-400 mV vs. Ag/AgCl.

It may be concluded that the use of dyadic modifiers leads to the appearance of the photoelectrochemical photocurrent switching effect in samples which did not exhibit it before (see the results for KCLA-containing materials, Figure S9) and its diminished manifestation in samples in which it was observed previously (see the spectra recorded for CLA-containing systems in Figures S10-S11). It means that GO mediates some kind of an additional charge transfer process which decreases the efficiency of the cathodic currents generation.

Similar results were obtained for oxygenated electrolyte solution and for the samples in which $TiO_2$ was modified with the use of CLA (Figures S10-S11). As expected, in the case of cadmium sulphide the impact of dyadic modifiers was much smaller, as the interaction between nanoparticles and molecular counterparts was less pronounced – probably due to the electronic configuration of cadmium ions on the surface (Figures S12-S13).

Based on the collected data we formulated a mechanism which could be responsible for these observations. In the proposed explanation graphene oxide plays a role of an efficient trap for electrons excited to the conduction band within the semiconductor. These trapping events are partially irreversible for potential values falling in the range between the $TiO_2$ conduction band edge and the bottom of the GO conduction band (approximately -0.45 – 0.0 V vs. Ag/AgCl) and fully irreversible at higher potentials. Thus, the further electron transport may occur exclusively between GO and the electrode or GO and quinone derivatives. The direct effect of such perturbation for the system is a decrease in the cathodic photocurrents intensity (as observed in the presented maps, Figure 8d).

Since no interactions between molecular entities were found, we believe that the surface of the semiconductor is a mediator between them. In the first case, when pure quinone derivatives are used as modifiers we end up with a very classical picture with partial sensitisation of the semiconductor and the excitation which occurs within nanoparticles and/or molecules and leads – depending on the electrode potential – to the generation of either anodic (Figure 9a) or cathodic (Figure 9b) photocurrents. In the latter case we do not observe the impact of molecular oxygen presence (Figures S9-S11) which may indicate that another redox couple may accept electrons from the conduction band and be responsible for closing the circuit. Good candidates for this role are molecules of quinone derivatives which are not strongly bounded (physisorbed) to the semiconductors surface.

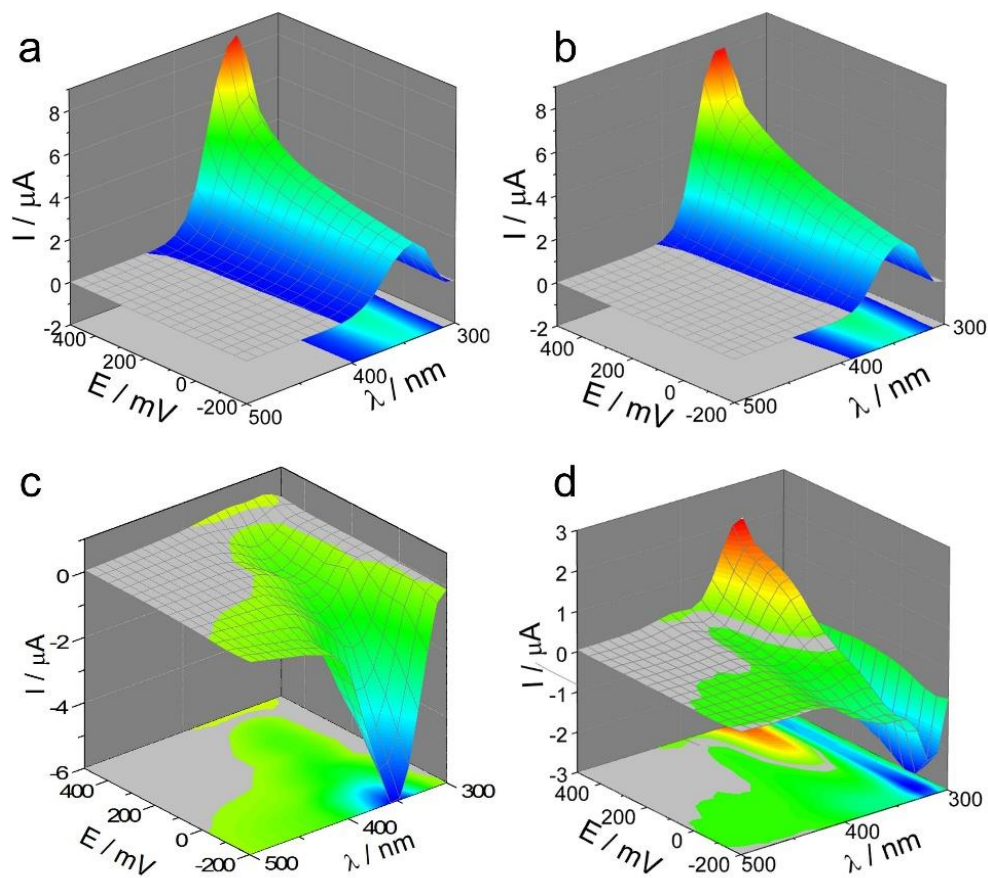

Figure 8. The photocurrent action spectra recorded at different potential steps for $TiO_2$ (a) modified with GO (b), KCLA (c) or KCLA/GO dyadic modifier (d) in 0.1 M $KNO_3$ under Ar. Potentials are given vs. Ag/AgCl (sat. KCl) electrode.

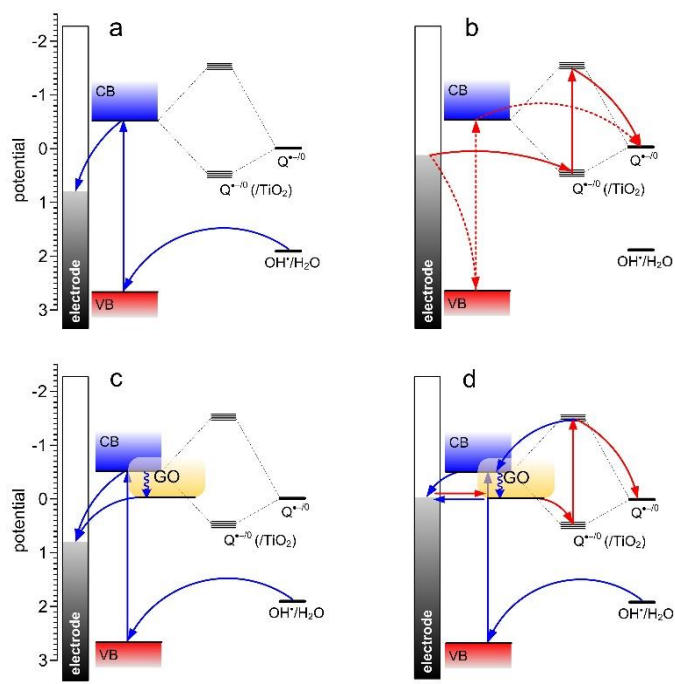

Figure 8. The schematic diagram describing the possible routes of charge carriers propagation within the investigated systems. It is important to note that quinone derivatives exist here in two forms – one which is strongly bound to the semiconductors surface (denoted as /$TiO_2$) and another, which is only weakly attached to $TiO_2$ nanoparticles. The anodic (A) and cathodic (B) photocurrent generation is shown. The influence of GO on the possible charge transfer pathways is explained in the case of the anodic photocurrents (C) and the cathodic ones (D). The grey bars in the left parts of the schemes represent the electrode potential.

Upon the addition of graphene oxide we noticed that the anodic signals are almost unaffected and rather enhanced than reduced. It may be a consequence of a hindered recombination in nanoparticles with the electron transfer occurring through GO states which mediates efficient anodic photocurrent generation (Figure 9c). Their position on the energy scale has been estimated based on the semiempirical calculations (Figure S14). Finally, in the case of the cathodic photocurrents generation in the material containing GO we observe a decrease in signal intensities and the photocurrent switching effect emerges due to the trapping events at the graphene oxide energy levels (Figure 9d). The excitation of strongly bound quinone derivatives is still possible but with GO acting as an electron sink a substantial number of charge carriers travel through GO electronic states and may be transferred towards the electrode is resonance processes (in the case the electrode potential reaches the GO states potentials range). Nonetheless, if the electrode potential is decreased above the upper (in the energy scale) edge of GO we may still observe relatively strong cathodic photocurrents as graphene oxide becomes semitransparent for electrons provided by the electrode – we simply recreate the situation from Figure 9b.

**Conclusions**

In the report, we presented an interesting example of dyadic modifiers used in hybrid materials with wide-bandgap semiconductors which exhibit no relevant interactions between molecular counterparts but strongly affect the photoelectrochemical properties of the semiconductor. We managed to find out that in the material made of titanium dioxide with chemisorbed KCLA/GO system some new pathways of charge transfer are possible – as a result of a strong binding of chloranilic acid to the $TiO_2$ surface new CT bands appear and some limited sensitization is observed. The profile of generated photocurrents also changes drastically as no anodic signals can be found for KCLA/$TiO_2$ hybrid.

Nonetheless, the most significant alteration is delivered by the introduction of graphene oxide. With it in the system we restore the ability of the materials to switch the photocurrents direction (so called PEPS effect) probably thanks to the additional trapping states which facilitate a directional transport of electrons excited to the conduction band to the electrode hindering, at the same time, the transfer in a cathodic mode. That observation may be essential for the design of photovoltaic devices consisting of carbon nanostructures and molecular sensitizers acting as efficient electron acceptors and may help in the construction of prototypic optoelectronic devices in which the current flow control at the molecular level is essential – e.g. they may be applied in the development of artificial synapses and neurons, as these biological structures relay on the directed signal flow.

**Acknowledgements**

Authors would like to thank Andrzej Blachecki for his assistance during measurements. This research has received funding from European Union's Horizon 2020 research and innovation programme under grant agreement No. 664786, the National Science Centre (grant no. UMO-2015/17/B/ST8/01783, UMO-2015/18/A/ST4/00058 and UMO 2013/11/D/ST5/03010), the Foundation for Polish Science (grant no. 71/UD/SKILLS/2014 carried-out within the INTER programme, co-financed from the European Union within the European Social Fund) and

Polish Ministry of Science and Higher Education within the "Iuventus Plus" programme in the years 2015-2017 (project no. 0256/IP2/2015/73). DFT calculations were performed at computing centre CYFRONET AGH within computational grant 'GRAPHENE'.